\documentclass[twocolumn,aps, showpacs]{revtex4}
\usepackage{graphicx}

\usepackage{amssymb}
\usepackage[latin1]{inputenc}
\usepackage{bm}

\newcommand{\nn}{\nonumber}
\newcommand{\Tr}{\mathrm{Tr}}
\newcommand{\mumax}{\mu_\mathrm{max}}

\begin{document}

\title{Effect of inelastic scattering on spin entanglement detection through current noise}
\author{Pablo San-Jose}
 \email{pablo@tfp.uni-karlsruhe.de}
 \affiliation {Institut für Theoretische Festkörperphysik, Universität Karlsruhe, 76128 Karlsruhe, Germany}
\author{Elsa Prada}
 \email{elsa@tfp.uni-karlsruhe.de}
 \affiliation {Institut für Theoretische Festkörperphysik, Universität Karlsruhe, 76128 Karlsruhe, Germany}

\date{\today}

\begin{abstract}
We study the effect of inelastic scattering on the spin entanglement
detection and discrimination scheme proposed by Egues, Burkard, and
Loss [Phys. Rev. Lett. \textbf{89}, 176401 (2002)]. The
finite-backscattering beam splitter geometry is supplemented by a
phenomenological model for inelastic scattering, the
charge-conserving voltage probe model, conveniently generalized to
deal with entangled states.  We find that the behavior of shot-noise
measurements in one of the outgoing leads remains an efficient way
to characterize the nature of the non-local spin correlations in the
incoming currents for an inelastic scattering probability up to
$50\%$. Higher order cumulants are analyzed, and are found to
contain no additional useful information on the spin correlations.
The technique we have developed is applicable to a wide range of
systems with voltage probes and spin correlations.
\end{abstract}


\maketitle

\section{Introduction}

Electron spin has various crucial properties that make it an ideal
candidate for a robust carrier of quantum entanglement in solid
state systems. Its typical relaxation and dephasing times can be
much larger than any other electronic timescale
\cite{Zutic04,Golovach04}, in particular in semiconductor
heterostructures, where its controlled manipulation begins to be a
reality \cite{Petta05}. This makes electron spin very valuable not
only in the context of spintronics \cite{Recher00}, but also in the
path to a scalable realization of a potential quantum computer.

Moreover, the possibility of demonstrating non-local quantum
entanglement of massive particles such as electrons is of conceptual
relevance in itself, since it is at the core of the quantum world
weirdness. Quantum optics are far ahead in this respect, and present
technology can already entangle \cite{Kwiat95}, teleport
\cite{Bouwmeester97} or otherwise manipulate quantum mechanically
\cite{Rauschenbeutel00} the polarization state of photons, and even
commercial solutions have been developed \cite{MagiQ} for completely
secure cryptographic key exchange via optical quantum communication.

In the context of solid state the equivalent feats are far away
still, due to the additional difficulties imposed mainly by the fact
that massive particles such as electrons suffer from interactions
with their environment, which can be in general avoided in the case
of photons. This in turn leads to strong decoherence effects, which
degrades the entanglement transportation. Sometimes these disruptive
effects can be minimized in the case of electron spin with the
proper techniques \cite{Petta05}. Still, the problem of controlled
spin manipulation and spin detection are two great hurdles to be
tackled in the long path to spin-based quantum computation
\cite{Loss98}. The main difficulty in the manipulation problem is
that all the operations available in usual electronics address
electron charge, being completely independent of the electron's
spin, unless some additional mechanism involving, e.g., external
magnetic fields \cite{Recher00, Hanson04}, ferromagnetic materials
\cite{Ohno98}, or spin-orbit coupling \cite{Koga02,Kato04b} are
relevant. Such mechanisms usually correlate spin states to charge
states, which allows to manipulate and detect the charge states via
more conventional means.

Several recent theoretical works have specifically studied the
influence of an electromagnetic environment
\cite{vanVelsen03,Samuelsson03Doga,Beenakker05} and the decoherence
through inelastic processes \cite{Prada05PRB,Taddei05} on orbital
and spin-entangled states, such as those that are the subject of the
present work. Generally, in all of these cases some type of spin
filter was necessary to measure the Bell inequalities, which makes
their experimental realization rather challenging.

Another interesting possibility to manipulate and detect spin states
with electrostatic voltages is through Pauli blocking, which appears
as a spin-dependent `repulsion' between two electrons due to Pauli
exclusion principle, as long as the two electrons share all the
remaining quantum numbers. This peculiarity is therefore specific of
fermions, and has no analog in quantum optics. An example of the
potential of such approach was illustrated in Ref.
\onlinecite{Burkard00}. It relied on the use of the mentioned Pauli
blocking mechanism in a perfect four-arm beam splitter supplemented
by the bunching (antibunching) behavior expected for symmetric
(antisymmetric) spatial two-electron wavefunctions. This was done
through the analysis of current noise \cite{Burkard00},
cross-correlators \cite{Samuelsson04}, and full counting statistics
(FCS) \cite{Taddei02}. It was also shown that it is possible to
distinguish between different incoming entangled states
\cite{Egues02,Samuelsson04}. In Ref. \onlinecite{Egues02} it was
demonstrated how the shot noise of (charge) current obtained in one
of the outgoing leads was enough to measure the precise entangled
state coming in through the two input arms, and to distinguish it
from a classical statistical mixture of spin states. Finite
backscattering and arbitrary mixtures in the spin sector were also
considered in Refs. \onlinecite{Burkard03} and \onlinecite{Egues05}.
Two channel leads and a microscopic description of the spin-orbit
interaction were also recently analyzed in great detail
\cite{Egues05}.

In this work we will analyze the robustness of the entanglement
detection scheme proposed in Ref. \onlinecite{Egues02} in the
presence of spin-conserving inelastic scattering and finite
beam-splitter backscattering for various entangled current states.
Although the spin sector is not modified by scattering, inelastic
scattering changes at least the energy quantum number of the
scattered electrons, and since Pauli exclusion principle does no
longer apply to electrons with different energy, we should expect
such inelastic processes to degrade the performance of the detection
scheme. From a complementary point of view, viewing the entangled
electron pairs as wavepackets localized in space, it is clear that
inelastic scattering will cause delays between them that will in
general make them arrive at the detectors at different times,
thereby lifting the Pauli blocking imposed by their spin
correlations \footnote{Note that this argument also applies to
elastic scattering as long as no energy filters are present before
the scatterer. Otherwise, mere elastic delay effects will be
irrelevant \cite{Samuelsson04}, and only inelastic scattering will
break Pauli blocking.}.

Moreover, as noted in Refs. \onlinecite{Burkard03} and
\onlinecite{Egues05}, the presence of backscattering introduces
spurious shot noise that is unrelated to the entanglement of the
source. Assuming known backscattering but, in general, unknown
inelastic scattering rate we show that the scheme remains valid in
certain range of parameter space, and point to a modified data
analysis to extract the maximum information out of local shot noise
measurements. We further study the information that may be extracted
from higher order cumulants of current fluctuations.

We will work within the scattering matrix formalism, and to describe
inelastic scattering we will employ a modification of the fictitious
voltage probe phenomenological model
\cite{Buttiker86,Buttiker88,Blanter00} generalized to include
instantaneous current conservation \cite{Beenakker92} in the
presence of spin correlated states. This approach relies on
phenomenological arguments and defines a scattering probability
$\alpha$ that is used to parametrize inelastic effects. Elastic
scattering has also been formulated within this language
\cite{deJong96}. The validity of the model has been widely
discussed, in general finding good qualitative agreement with
microscopic models \cite{Kiesslich05,Brouwer97,FoaTorres05,Texier00}
and experiments \cite{Oberholzer05}. Recently it was demonstrated to
become equivalent to microscopic phase averaging techniques at the
FCS level in some limits and setups \cite{Pilgram05} (clarifying
some apparent discrepancies with classical arguments
\cite{Marquardt04}). Also recently, it has been applied to study the
effect of spin relaxation and decoherence in elastic transport in
chaotic quantum dots \cite{Michaelis05,Beenakker06}. The scheme
remains attractive as a first approximation to inelastic (or
elastic) processes. Alternatively, it is a good model for a real
infinite-impedance voltage probe, a common component of many
mesoscopic devices. The generalization we present here is
specifically targeted towards the computation of the FCS of
mesoscopic systems with inelastic scattering and incoming scattering
states with arbitrary entanglement properties. The problem of how to
apply such decoherence model to the particularly interesting case of
non-locally entangled input currents has not been previously
discussed to the best of our knowledge, except in Ref.
\onlinecite{Prada05PRB}, where current conservation was not taken
fully into account.

This paper is organized as follows. In Sec. \ref{sec:system} we
discuss the beam-splitter device as an entanglement detector in the
presence of inelastic scattering. In Sec. \ref{sec:technique} we
give a short account of the technique we will employ to compute the
FCS. Further details on our implementation of the fictitious probe
scheme can be found in Appendix \ref{sec:ap}. A second Appendix
\ref{ap:comp} clarifies the connection between the Langevin approach
and the employed technique in a simple setup, and also illustrates
to what extent it succeeds or fails when it spin-correlations are
introduced. The analysis of the obtained results for the operation
of the device are explained in Sec. \ref{sec:results}. A summarized
conclusion is given in Sec. \ref{sec:conclusions}.

\section{Beam-splitter device with inelastic scattering \label{sec:system}}

\begin{figure}
\includegraphics[width=7cm]{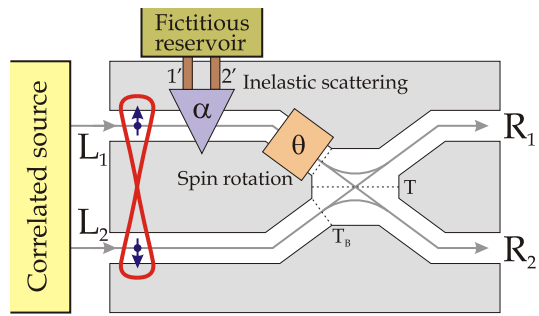}
\caption{(Color online) The beam splitter geometry fed with pairwise
non-locally entangled electron currents or polarized currents. The
action of the spin rotation via Rashba spin-orbit coupling in one of
the input leads changes noise in the output leads dramatically.
Inelastic transport is modeled between the entangler and the spin
rotation region by means of one or more fictitious probes. Shot
noise measured in terminal $R_1$ as a function of $\theta$ can be
used to detect the nature of the incoming electron correlations.}
\label{fig:setup}
\end{figure}

The system we will study is depicted in Fig. \ref{fig:setup}. It is
an electronic beam splitter patterned on a two-dimensional electron
gas (2DEG) with two (equal length) incoming and two outgoing arms,
such that the transmission probability between the upper and the
lower arms is $T$. The beam splitter is assumed to have also a
finite backscattering amplitude whereby electrons get reflected back
into the left leads with probability $1-T_B$. We have considered two
possibilities for backscattering: the technically simpler case
without cross reflection, for which electrons scatter back always
into their original incoming leads, and which we will term
\emph{simple} backscattering; and the fully symmetric case, whereby
the probability of going from any upper lead to any lower lead
remains $T$, be it on the left or the right, which we will call
\emph{symmetric} backscattering. This distinction is only relevant
when there is a finite inelastic scattering on the leads, and both
give very similar results in any case, so we will focus mainly in
the simple backscattering case \footnote{Admittedly, in a realistic
system there could be a finite probability that a backscattered
particle be scattered back onto the system, which would probably
give further corrections. We neglect these contributions for
simplicity, although they could easily be included (in the limit of
small induced delay) into the total scattering matrix by assuming a
contact between the entangler and the system of finite
transparency.}. Other authors \cite{Egues05} have previously studied
the effect of backscattering in this geometry, although considering
that only the electrons in the lead with the backgates can
backscatter, whereas in our case the two incoming leads are
equivalent (the scattering occurs in the beamsplitter). The effect,
as we shall see, is however qualitatively equivalent to their
result, which is that backscattering effectively reduces the
oscillation amplitude of noise with the spin rotation angle.

We connect the right arms to ground and the two incoming arms to a
reservoir that emits non-local spin-correlated electron pairs,
biased at a voltage $-V$. For definiteness we choose these pairs so
that the $\hat{z}$ spin component of the electron coming at a given
time through lead $L_1$  is always opposite to that of the
corresponding electron coming simultaneously through lead $L_2$.
They could be or not be entangled, depending on the characteristics
of the source and the leads from source to splitter. Time
coincidence of pairs is assumed to within a timescale $\tau_\Delta$
that is shorter than any other timescale in the system, such as
$\Delta t\equiv h/eV$. This implies two constraints. On the one
hand, if the source is an entangler such as e.g. that of Refs.
\onlinecite{Recher01,Prada04,Prada05NJP}, this would mean that the
superconductor emitting the correlated pairs has a large gap
$\Delta$ as compared to the bias voltage. On the other hand, the
length of the leads connecting the entangler to the beam-splitter
device should be of equal length to within $v_F\Delta t$ accuracy.

A local spin-rotation in lead $L_1$ is implemented by the addition
of backgates above and below a section of lead $L_1$. Applying a
voltage across these backgates the structure inversion asymmetry of
the 2DEG is enhanced, inducing a strong Rashba spin-orbit coupling
in that region of the 2DEG in a tunable fashion without changing the
electron concentration \cite{Miller03}. This in turn gives rise to a
precession of the spin around an in-plane axis perpendicular to the
electron momentum, which we chose as the $\hat{y}$ axis, resulting
in a tunable spin rotation of an angle $\theta$ around $\hat{y}$
after crossing the region with backgates.

The idea behind this setup is that the spin rotation can change the
symmetry of the spatial part of the electron pair wavefunction, thus
affecting the expected shot noise in the outgoing leads, which is
enhanced for even and suppressed for odd spatial wavefunctions. The
switching from bunching to antibunching signatures in the shot noise
as a function of $\theta$ is enough to identify truly entangled
singlets in the incoming current. Likewise, a $\theta$ independent
shot noise is an unambiguous signal of a triplet incoming current,
since a local rotation of a triplet yields a superposition of
triplets, preserving odd spatial symmetry and therefore,
antibunching. A current of statistically mixed anticorrelated
electron spins can also be distinguished from the entangled cases
from the amplitude of the shot noise oscillations with $\theta$.
Thus, this device was proposed as a realizable entanglement detector
through local shot noise measurements
\cite{Egues02,Burkard03,Egues05}.

As discussed in the introduction, inelastic scattering due to
environmental fluctuations could spoil the physical mechanism
underlying this detector, which is Pauli exclusion principle, and
should therefore be expected to affect its performance in some way.
The implementation of inelastic scattering in ballistic electron
systems can be tackled quite simply on a phenomenological level
through the addition of fictitious reservoirs within the scattering
matrix formalism \cite{Blanter00}. The necessary generalization to
deal with entangled currents and a simple scheme to derive the FCS
in generic systems with additional fictitious probes is presented in
appendix \ref{sec:ap}. We model spin-conserving inelastic scattering
by the addition of two fictitious probes (one for spin-up and
another for spin-down) in lead $L_1$, depicted as a single one in
Fig. \ref{fig:setup}. We have numerically checked that the addition
of another two fictitious probes in lead $L_2$ gives very similar
results for the shot noise through the system, so we will take only
two in the upper arm for simplicity. This is also physically
reasonable if we consider only decoherence due to the backgates
deposited on the upper arm to perform the local Rashba
spin-rotation, which provide a large bath of external fluctuations
that can cause a much more effective inelastic scattering. The
parameter that controls the inelastic scattering probability is
$\alpha\in[0,1]$, being $\alpha=1$ the completely incoherent limit.

In the following analysis we will inject into the input arms of the
device currents with different types of initial non-local electron-pair density matrix,
\begin{eqnarray}
\hat{\rho}=\frac{1}{2}\left(|L_1\!\!\uparrow;L_2\!\!\downarrow\rangle\langle
L_1\!\!\uparrow;L_2\!\!\downarrow\!|+|L_1\!\!\downarrow;L_2\!\!\uparrow\rangle\langle
L_1\!\!\downarrow;L_2\!\!\uparrow\!|\right)\nn\\
+\frac{\beta}{2}\left(|L_1\!\!\downarrow;L_2\!\!\uparrow\rangle\langle
L_1\!\!\uparrow;L_2\!\!\downarrow\!|+|L_1\!\!\uparrow;L_2\!\!\downarrow\rangle\langle
L_1\!\!\downarrow;L_2\!\!\uparrow\!|\right),
\label{states}
\end{eqnarray}
namely, (i) statistical mixtures of up and down classically
correlated electrons (diagonal density matrix, $\beta=0$), which we
will also call spin-polarized currents, (ii) EPR-type singlet
spin-entangled pure states ($\beta=-1$), and (iii) idem with $m_s=0$
triplet states ($\beta=1$). We will use subindexes $s$, $t$, and $m$
to denote the pure singlet, pure $m_s=0$ triplet and statistically
mixed incoming states. Note that this expression refers to pairs of
electrons that arrive at the same time at the device, so that this
density matrix is actually expressed in a localized wavepacket
basis.

Our goal is to ascertain to what extent, for a splitter transmission
$T$, a finite backscattering $1-T_B$ and finite and unknown amount
of inelastic scattering $\alpha$ in the input leads, the shot noise
in one of the output arms ($R_1$) as a function of rotation angle
$\theta$ could still be used to demonstrate the existence or not of
initial entanglement, and that way provide a means to distinguish
truly quantum-correlated states from statistically correlated
(unentangled) ones.

\section{The technique\label{sec:technique}}

In Appendix \ref{sec:ap} we give a detailed account of the method we
have used, which can be employed to compute the FCS of a generic
mesoscopic conductor with instantaneous current conservation (on the
scale of the measuring time) in the attached voltage probes, and
generic spin correlations in the incoming currents. We work within
the wave packet representation, whereby the basis for electron
states is a set of localized in space wavefunctions \cite{Martin92}.
A sequential scattering approximation is implicit, which however
yields the correct $\omega=0$ current fluctuations in known cases
with inelastic scattering, see, e.g., Appendix \ref{ap:comp}. We
summarize here the main points as a general recipe for practical
calculations.

Given a certain mesoscopic system with a number of biased external
leads connected to reservoirs, one should add the desired voltage
probes to model inelastic scattering (or real probes), and perform
the following steps to compute the long-time FCS of the system:

(i) Define the (possibly entangled) incoming states in the external
leads for a single scattering event without the probes,
\begin{equation}
|in\rangle=R[\{\hat{a}^+\}] |vac\rangle.
\end{equation}
Here $R[\{\hat{a}^+\}]$ is an arbitrary combination of creation
operators $a^+_n$ of incoming electrons (in the localized wavepacket
basis) acting on the system's vacuum. In our case it would create
state (\ref{states}).

(ii) Add the $N$ two-legged voltage probes (one channel per leg)
with individual scattering matrices as in Eq. (\ref{Salpha}), and
compute the total $S$-matrix of the multi-terminal system, $S_{nm}$.
Note that $S_{nm}(t,t')$ in our temporal basis is assumed to be
constant, i.e., independent of $t,t'$, which corresponds to an
energy independent scattering matrix in an energy basis.

(iii) Define outgoing electron operators $\hat{b}^+_n=\sum_m
S_{nm}\hat{a}^+_m$. To implement instantaneous current conservation
we expand our Hilbert space with $N$ integer slave degrees of
freedom $\vec{Q}=\{Q_i\}$, which result in the following outgoing
state after one scattering event,
\begin{equation}
|out;\vec{Q}\rangle \equiv
R[\{\hat{b}^+\}]\prod_{i}^N\left[\hat{b}^+_{p_i;1'}\hat{b}^+_{p_i;2'}\right]^{g(Q_i)/2}|vac\rangle.
\end{equation}
These $Q_i$ are counters of total charge accumulated in the probes.
The notation here is that $\hat{b}^+_{p_i;l}$ creates the scattered
state resulting from an electron injected through leg $l=\{1',2'\}$
of the two-legged probe $i$. $g(Q)$  encodes the response of the
probe to a certain accumulated charge $Q$. The specific form of
$g(Q)$ is not essential as long as it tends to compensate for any
charge imbalance in the probe. One convenient choice is given in Eq.
(\ref{choice}), which yields in our setup a minimal tripled-valued
fluctuation interval of $Q_i\in[-1,1]$. Note also that state
$|out;\vec{Q}\rangle$ in the above equation is nothing but
$U_{\Delta t}|\phi^e_j\vec{Q}\rangle$ of Appendix \ref{sec:ap}.

(iv) Compute the $3^N\times 3^N$  $\overline{W}$ matrix
\begin{equation}
\overline{W}_{\vec{Q}_b\vec{Q}_a}(\vec{\lambda})=\langle
out;\vec{Q}_a|\hat{P}_{\vec{Q}_b}\hat{\chi}_j(\vec{\lambda})
|out;\vec{Q}_a\rangle, \label{Wbar2}
\end{equation}
which we write in terms of the moment generating operator
$\hat{\chi}_j=e^{i\sum_n
\lambda_n(\hat{N}^\mathrm{out}_{n}-\hat{N}^\mathrm{in}_{n})}$, where
$\hat{N}$ are the number operators of electrons scattering in event
$j$. The operator $\hat{P}_{\vec{Q}_b}$ projects onto the subspace
of electron states that have a total of ${Q_b}_i-g({Q_a}_i)$
particles scattered into probe $i$, i.e., states in which the probe
$i$ has gone from ${Q_a}_i$ to ${Q_b}_i$ excess electrons. If the
incoming state is not a pure state, one should perform the
statistical averaging over the relevant $| out;\vec{Q}\rangle$
states at this point.

(v) Compute the resulting long-time \emph{current} moment generating
function $\chi_I(\vec{\lambda})$ by taking the maximum eigenvalue of
matrix $\overline{W}$. The charge generating function
$\chi(\vec{\lambda})$ is obtained simply by taking the power $M$ of
$\chi_I(\vec{\lambda})$, cf. Eq. (\ref{ChiIDef}), where $M=eVt/h$ is
the average number of emitted pairs from the source after an
experiment time $t$ at a bias $V$.

We make use of this method in our particular system by setting a
single counting field $\lambda$ on output lead $R1$, where we wish
to compute current fluctuations. This way we derive results for
$\chi_I$ and current cumulants [see Eqs. (\ref{Icum}) and
(\ref{chiI})] from the corresponding $\overline{W}$ matrix
(\ref{Wbar2}) for the different types of injected currents of Eq.
(\ref{states}).

While explicit expressions for the current cumulant generating
function $\ln\chi_I(\lambda)$ are in general impossible due to the
large dimensions of the $\overline{W}$ matrix ($9\times 9$ in this
case), it is always possible to write $\chi_I$ in an implicit form
that is just as useful to sequentially compute all cumulants,
namely, the eigenvalue equation
\begin{equation}
\det\left[\overline{W}(\lambda)-\chi_I(\lambda)\bm{1}\right]=0,
\label{chidet}
\end{equation}
supplemented by the condition $\chi_I(0)=1$. By differentiating this
equation around $\lambda=0$ a number of times and using
(\ref{Icum}), one can obtain the various zero-frequency current
cumulants on arm $R1$.

In the next section, instead of giving the general expression of
$\overline{W}$, which is rather large, we provide the explicit
expressions for $\chi_I$ and shot noise obtained in various useful
limiting cases, together with plots of the first cumulants in the
$\{T,T_B,\alpha,\theta\}$ parameter space.

\section{Results \label{sec:results}}

In this section we will analyze the performance of the beam splitter
device of Fig. \ref{fig:setup} as a detector of quantum correlations
in the incoming currents through the shot noise or higher current
cumulants induced in arm $R_1$. We will first make connection with
the results in the literature \cite{Egues02} by computing the shot
noise in an elastic splitter, and then we will generalize them to
finite inelastic scattering probabilities and finite backscattering.
We will thus establish tolerance bounds for such imperfections in
the detector. Finally, we will address the question of whether the
measurement of higher order current cumulants could improve the
tolerance bounds of the device.

\subsection{Shot noise}

\begin{figure}
\includegraphics[width=7cm]{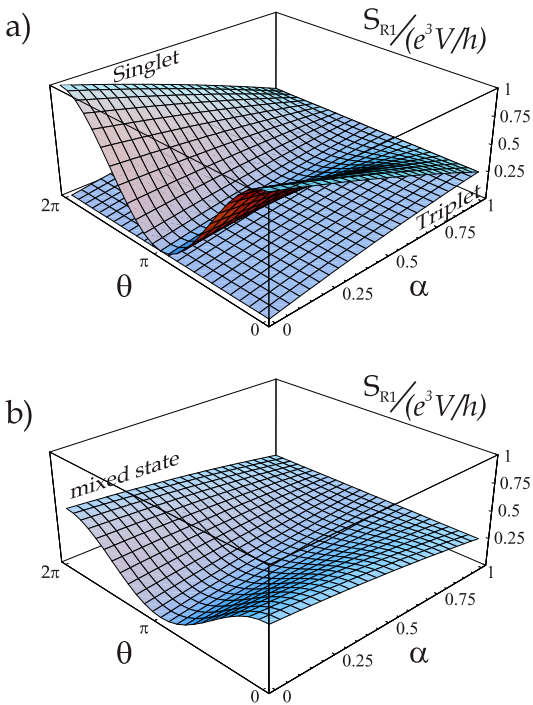}
\caption{(Color online) In the upper plot (a) we represent the
current shot-noise in units of $e^3 V/h$ in lead $R1$ for the
singlet and $m_s=0$ triplet incoming states, as a function of spin
rotation angle $\theta$ and decoherence strength $\alpha$. The same
for the polarized spin state case is presented in the lower plot
(b). Inter-lead transmission probability between upper and lower
arms $T$ is fixed to $0.5$, and no backscattering ($T_B=1$) is
assumed.} \label{fig:3d}
\end{figure}

In the elastic transport limit $\alpha=0$ and with arbitrary
intralead backscattering strength $1-T_B$, the following expression
for the shot noise is obtained,
\begin{equation}
S=\frac{e^3 |V|}{h}T_B\left[1-T_B+(1-\beta) T_B
T(1-T)(1+\cos\theta)\right], \label{S(alpha=0)}
\end{equation}
where constant $\beta$ corresponds to the different types of
incoming current, cf. Eq. (\ref{states}). Note that this expression
holds for simple or symmetric backscattering (as defined in Sec.
\ref{sec:system}). As shown by Eq. (\ref{S(alpha=0)}), for
$\alpha=0$ the amplitude of the $\theta$ dependence is enough to
distinguish between the different types of states, if $T$ and $T_B$
are known. As could have been expected, the triplet current noise
($\beta=1$) is $\theta$ independent, since the local spin rotation
only transforms the $m_s=0$ triplet to a different superposition of
the other triplet states, none of which can contribute to noise
since each electron can only scatter into different outgoing leads
due to the Pauli exclusion principle.

However, in the presence of a strong coupling to the environment,
$\alpha=1$, the shot noise behaves very differently. Due to the
complete incoherence of scattering, which changes the orbital
quantum numbers (or arrival times at the detector) of the incoming
states, the bunching-antibunching switching disappears. Therefore
$S_s$, $S_t$ and $S_m$ become equal and $\theta$ independent. In
particular, for simple backscattering we have
\begin{eqnarray}
S_s=&S_t&=S_m=\frac{e^3 |V|}{h}
\frac{T_B}{(1+T_B)^3}\left[1+2T_B^2T(1-T)
\right.\nn\\
&&\left.-T_B(1+3T^2-6T)-T_B^3T^2(2+T_B)\right]. \label{S(alpha=1)}
\end{eqnarray}

These features are illustrated in Fig. \ref{fig:3d}, where we have
plotted the current shot-noise in lead $R_1$, normalized to the
constant $e^3 V/h$, \footnote{The shot noise normalized to $e^3 V/h$
is in fact the Fano factor since the total current is $I_{R_1}=e^2
|V|T_B/h$.} as a function of the spin rotation angle $\theta$ and
the decoherence parameter $\alpha$, at $T_B=1$ and $T=1/2$:
\begin{eqnarray}
\frac{S_s}{e^3 V/h}&=&\frac{1}{2}\left[1-(\alpha-1)\cos\theta\right]-\frac{\alpha}{32}(5+3\alpha),\\
\frac{S_t}{e^3 V/h}&=&\frac{\alpha}{32}(11-3\alpha),\\
\frac{S_m}{e^3
V/h}&=&\frac{1}{4}\left[1-(\alpha-1)\cos\theta\right]-\frac{\alpha}{32}(3-3\alpha).
\end{eqnarray}
Note that at $\theta=\pi$ these three shot noise curves are all
equal. Note furthermore that for $\alpha=1$ we have
$S_s=S_t=S_m=(e^3 V/h) T(1-T)$ which is $1/4$ in normalized units.
The cosine-type dependence of the current noise with $\theta$,
$S(\theta)=S(\pi)+\Delta S\cos^2(\theta/2)$, where $\Delta
S=S(\theta=0)-S(\theta=\pi)$, holds for any value of $\alpha<1$ in
the singlet and polarized cases. The oscillation amplitude of the
noise for the singlet case is always twice the oscillation amplitude
of the polarized one. In contrast, the triplet shot noise (and all
higher cumulants for that matter) remains always $\theta$
independent for any $\alpha$ and $T_B$.

Since our aim in this study is to find a way to distinguish between
the different incoming states of Eq. (\ref{states}), we will
disregard from now on the trivial case of the triplet current, which
is easily detectable by its $\theta$-independence, and focus
entirely on the distinction between the singlet and mixed state
cases. In these two cases, when $T_B<1$, the oscillatory behavior
with $\theta$ remains, although it is no longer purely sinusoidal.
Besides, its oscillation amplitude quickly decreases with increasing
backscattering, making the entanglement detection scheme harder.
However, we will now show that, knowing only the value of the
shot-noise at zero spin rotation angle (or alternatively the
amplitude $\Delta S$), it is possible to distinguish between the
different incoming states for not-too-strong decoherence.

\subsection{Robust entanglement detection scheme}

\begin{figure}
\includegraphics[width=8cm]{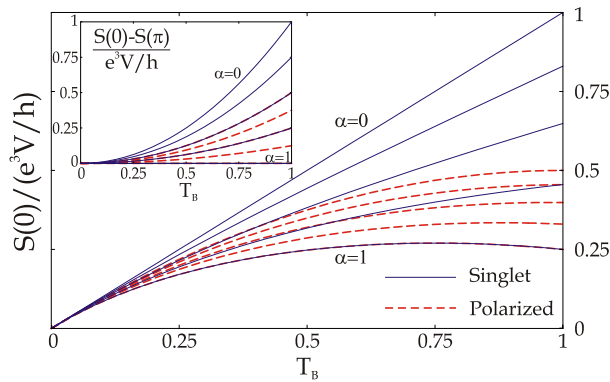}
\caption{(Color online) Normalized value of shot-noise in lead $R_1$
at zero spin rotation angle as a function of beam splitter
transmission $T_B$ for $T=0.5$. Solid (blue) lines correspond to
singlet incoming current whereas dashed (red) lines account for the
polarized one. In both cases, different values of inelastic
scattering probability have been considered, from $\alpha=0$ (upper
curves) to $\alpha=1$ (lower curves) in steps of $0.25$. Inset: the
same for the oscillation amplitude with $\theta$ of the shot-noise.}
\label{fig:noise}
\end{figure}

Tuning once again the beam splitter to the symmetric $T=1/2$ point,
which turns out to be the optimum point of operation for
entanglement detection, we notice from Fig. \ref{fig:3d} that the
analysis of the $\theta$ dependence of the shot-noise at an
arbitrary and unknown value of $\alpha$ indeed precludes from a
clear distinction of the singlet and mixed state cases.

A more complete picture can be obtained by plotting the value of the
shot-noise at $\theta=0$ in the interval $\alpha\in[0,1]$ as a
function of $T_B$. This is done in Fig. \ref{fig:noise} for the case
of simple backscattering. Solid (blue) lines correspond to singlet
incoming current, and dashed (red) lines to the polarized one.
Moreover, the upper curve in both sets of curves accounts for the
case of $\alpha=0$, and for the next ones the value of the inelastic
scattering parameter increases, in steps of $0.25$, until $\alpha=1$
for the lower curves (which coincide for both the entangled and
polarized cases). The same analysis can be done for the behavior of
the amplitude $\Delta S$ as a function of $T_B$, as shown in the
inset of Fig. \ref{fig:noise} (given also for simple
backscattering). In this latter case, the amplitudes for both the
singlet and the polarized currents have in fact a very simple
analytical form, the singlet case ranging from $T_B^2$ to $0$ and
the polarized one from $T_B^2/2$ to $0$ as we sweep from $\alpha=0$
to $\alpha=1$. Therefore, we see how the $\theta$-independent
background noise introduced by the finite backscattering in the main
plot of Fig. \ref{fig:noise}, which could in principle degrade the
performance of the entanglement detector as mentioned in Ref.
\onlinecite{Burkard03}, can be filtered out by measuring the
amplitude $\Delta S$. We also note that if a symmetric
backscattering is considered, the resulting curves for Fig.
\ref{fig:noise} are qualitatively the same, and therefore it does
not affect the above discussion.

We can observe in both plots of Fig. \ref{fig:noise} that if
$\alpha$ is unknown, as it is usually the case in an experiment, the
classical and quantum currents are distinguishable from a single
noise measurement (or two in the case of the inset) only if its
value is found to lie outside of the overlapping region between the
two sets of curves. According to this model, this should always
happen for values of inelastic scattering smaller than at least one
half. In the case of the main figure, even higher values of $\alpha$
can be distinguished for values of $T_B$ close to one. In any case,
the values of $\alpha$ for which the noise measurement is no longer
able to distinguish a singlet entanglement from a statistically
mixed case are rather high, $\alpha\in[0.5,1]$. This means that, in
a realistic situation where decoherence is not too strong,
shot-noise measurements remain enough for determining if the source
feeding the beam-splitter is emitting entangled or statistically
mixed states.

\subsection{Higher order cumulants}

\begin{figure}
\includegraphics[width=7cm]{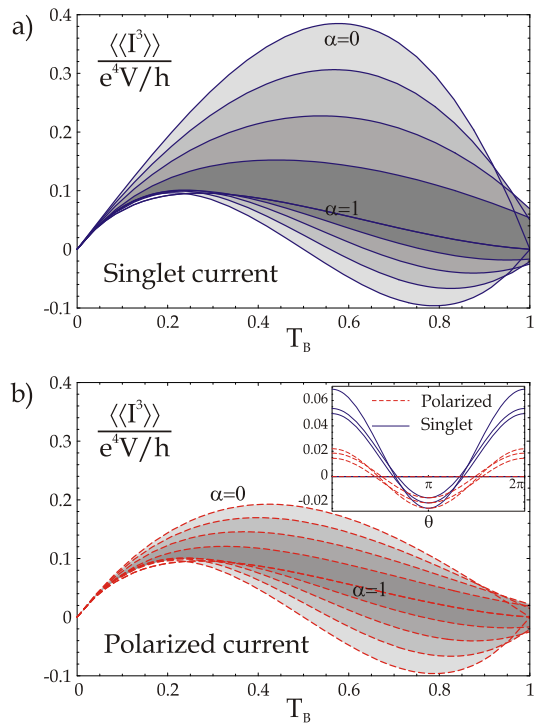}
\caption{(Color online) Oscillation range with angle $\theta$ of the
skewness in lead $R_1$, in units of $e^4 V/h$, as a function of
beam-splitter transmission between left and right arms $T_B$. $T$ is
fixed to the optimal point $T=0.5$. Different oscillation ranges for
different values of inelastic scattering are indicated by different
shades of gray, ranging from pale gray for $\alpha=0$ to dark gray
for $\alpha=1$ in steps of 0.25. The case of singlet-entangled
incoming current is considered in plot (a), whereas in plot (b) the
incoming current is in a polarized state. The actual oscillation of
skewness with Rashba spin rotation angle for entangled current
(solid blue lines) and polarized current (dash red lines) is plotted
in the inset of (b) (see main text).} \label{fig:skewness}
\end{figure}

We could ask whether it is possible to distinguish between incoming
singlet-entangled and polarized currents for a wider range of
parameters $\alpha$ by analyzing higher order cumulants. The short
answer is "no".

As we did for the noise in Fig. \ref{fig:3d}, we can plot the
angular dependence of the third moment, the skewness, for different
values of inelastic scattering parameter $\alpha$. This is shown in
a 2D plot in the inset of Fig. \ref{fig:skewness}(b) for $T=0.5$ and
$T_B=1$. As before, solid (blue) lines and dashed (red) lines
correspond to spin singlet-entangled and polarized incoming
currents, respectively. Now we find that the behavior of skewness
with $\theta$ is not monotonous as $\alpha$ varies. For $\alpha=0$
and $\alpha=1$ at $T_B=1$ the third cumulant is zero for every angle
both for entangled and for mixed states (the probability
distribution of the current is symmetric for those parameters as was
previously noted in the $\alpha=0$ case in Ref. \cite{Taddei02}).
Moreover, this means that the skewness is not a good entanglement
detector for a near perfect beam splitter, nor when the inelastic
scattering is strong. For intermediate values of decoherence, still
at $T_B=1$, the skewness oscillates with the spin rotation angle and
its oscillation amplitude, $\langle\langle I^3
\rangle\rangle(\theta=0)-\langle\langle I^3
\rangle\rangle(\theta=\pi)$, has a maximum around $\alpha\approx
0.5$. This oscillation range is depicted in the main plots of Fig.
\ref{fig:skewness} as a function of the transmission between left
and right arms $T_B$ (where simple backscattering has been
considered). Several values of the inelastic parameter are
differentiated using different shades of gray, ranging from pale
gray for $\alpha=0$ to dark gray for $\alpha=1$ in steps of 0.25.
The main features can be summarized as follows. First, both for the
entangled and the polarized current, the broadest oscillation range
occurs for $\alpha=0$ (being bigger for the singlet-entangled case).
Second, for $\alpha=1$ the oscillation amplitude in both cases is
zero, although the skewness remains finite and positive (shifting
from a Gaussian to a Poissonian distribution of current as $T_B$
goes from $1$ to $0$). For $T_B$ smaller than $0.9$ approximately,
the behavior of the oscillation range is monotonous with $\alpha$,
it simply decreases with it. For small values of $T_B$, the skewness
coincides with the shot-noise, which is expected since the
probability distribution for a tunnel barrier recovers a Poisson
distribution, even in the presence of inelastic scattering. In
general, the sign of the skewness is reversed in a wide range of
parameters by tuning the spin rotation $\theta$.

Concerning our entanglement detection motivation, comparing Fig.
\ref{fig:skewness}(a) and \ref{fig:skewness}(b) we find that the
third cumulant does not provide any further information in our
search of a way to distinguish between entangled and non-entangled
incoming states. For $T_B$ of the order of $0.9$ and above, due to
the non-monotonous behavior of the oscillation range with $\alpha$,
the skewness of our beam splitter setup can hardly be used as a
detector of entanglement at all. For smaller values of $T_B$, we are
able to discriminate between different currents in the same range of
inelastic scattering parameter $\alpha$ as we could with the noise,
this is, from zero decoherence to roughly $\alpha=0.5$.

We have also analyzed further cumulants, whose behaviors with $T_B$
and $\theta$ get more intricate as the order of the cumulant
increases, and have found the same qualitative result. Either they
are not useful tools for entanglement detection or the range of
parameters $\alpha$ and $T_B$ is not improved from what we find with
the shot-noise measurements.

\section{Conclusions \label{sec:conclusions}}

In this work we have analyzed the effect of inelastic scattering,
modeled by spin-current conserving voltage probes, on entanglement
detection through a beam-splitter geometry. We have shown that the
action of inelastic processes in the beam-splitter cannot be
neglected, since it directly affects the underlying physical
mechanism of the detector, which is the fact that two electrons with
equal quantum numbers cannot be scattered into the same quantum
channel. If there is a finite inelastic scattering, such
antibunching mechanism is no longer perfect, and the entanglement
detection scheme has to be revised.

However, we have found that detection of entanglement through shot
noise measurements remains possible even under very relaxed
conditions for imperfections in the beam-splitter device and
substantial inelastic scattering. Even if a reliable microscopic
description of inelastic processes is not available, the present
analysis suggests that the detection scheme is robust for inelastic
scattering probabilities up to $50\%$.

We have also shown that higher current cumulants do not contain more
information about the entanglement of the incoming currents than the
shot noise. We have analyzed in particular the skewness of current
fluctuations, finding that finite backscattering and inelastic
scattering strongly affect the asymmetry of current fluctuations. In
particular, a positive skewness is developed as the beam splitter
transparency is lowered.

Finally, we have developed a novel way to implement current
conservation in voltage probe setups when the incoming currents are
non-locally spin correlated, which can be applied to a wide variety
of problems where entanglement is key.

\begin{acknowledgments}
The authors would like to acknowledge the inspiring conversation
with M. Büttiker that encouraged us to develop our formulation of
the voltage probe FCS with general spin correlations. The authors
also enjoyed fruitful discussions with D. Bagrets, F. Taddei, R.
Fazio and C.W.J. Beenakker. This work benefited from the financial
support of the European Community under the Marie Curie Research
Training Networks, ESR program, the SQUBIT2 project with number
IST-2001-39083 and the MEC (Spain) under Grants Nos. BFM2001-0172
and FIS2004-05120.
\end{acknowledgments}

\appendix

\section{Phenomenological description of inelastic scattering \label{sec:ap}}

Voltage probes are frequently real components of mesoscopic devices,
but have also been used traditionally for phenomenological modeling
purposes. The voltage probe description of inelastic scattering
resorts to the addition of one or more fictitious reservoirs and
leads attached to the coherent conductor under study through
specific scattering matrices, around the regions where inelastic
scattering is to be modeled. While being still coherent overall, the
elimination of the fictitious reservoirs results in an effective
description of transport such that electrons that originally
scattered into the reservoirs now appear as having lost phase and
energy memory completely.

We will now discuss the implementation of the voltage probe in the
presence of charge relaxation and general incoming states. The whole
idea of the voltage probe is to use the non-interacting scattering
formalism to model inelastic electron scattering, and the crossover
from coherent conductors to incoherent ones. There are two ways to
do this. The simpler one assumes a static chemical potential in the
probes that is computed self-consistently by fixing time-averaged
current flowing into the probes to zero, as corresponds to an
infinite impedance voltage probe, or to inelastic scattering. This
gives a physically sound conductance value, but fails to yield
reasonable shot noise predictions. The reason is that total current
throughout the system should be instantaneously conserved. The more
elaborate way, therefore, assumes fluctuations in the state of the
probe that can compensate the current flowing into the probe(s) at
any instant of time (and possibly also energy if one is modeling
pure elastic dephasing \cite{deJong96,vanLangen97,Chung05}), which
gives results for current fluctuations in agreement with classical
arguments \cite{Blanter00}.

It is traditional to impose such constraint within a Langevin
description of current fluctuations \cite{Beenakker92}, whereby the
chemical potential in the probe is allowed to fluctuate, but the
intrinsic formulation of this approach makes it inadequate to treat
the statistics of general incoming (entangled or mixed) states,
other than those produced by controlling individual chemical
potentials. There seems to be no known way of how to include the
effect of arbitrary correlations between electrons, such as e.g.
non-local quantum spin correlations, which are precisely the
contributions we are interested in this work. The technique we will
develop in the following explicitly takes into account the precise
incoming state of the electrons, and recovers results obtained
within the Langevin approach in the case of non-correlated incoming
states. For an analysis of the possibility of implementing these
correlation effects within the Langevin approach and for a
comparison to the present (time-resolved) technique, see Appendix
\ref{ap:comp})

The scattering matrix to a (two-legged \footnote{If the channels
were chiral, i.e. no backscattering allowed, one could take
single-legged probes as well. For non-chiral channels it is in
general impossible to model full decoherence without introducing at
least two legs.}) fictitious probe is given, in the basis
${1,2,1',2'}$ (being $1',2'$ the extra leads), by
\begin{equation}
S_\alpha=\left(\begin{array}{cccc}
0               &-\sqrt{1-\alpha}   &i\sqrt{\alpha}     &0                 \\
\sqrt{1-\alpha} &0                  &0                  &i\sqrt{\alpha}           \\
i\sqrt{\alpha}  &0                  &0                  &\sqrt{1-\alpha}   \\
0               &i\sqrt{\alpha}     &-\sqrt{1-\alpha}   &0
\end{array}\right),
\label{Salpha}
\end{equation}
with $\alpha$ being the inelastic scattering probability. This
should be composed together with any other scattering matrices in
the system, and any other probes present. In a spinful case in which
inelastic scattering does not flip spin there should be at least two
of these probes, one per spin channel. Other considerations such as
inelastic channel mixing in multichannel cases should be taken into
account when designing the relevant fictitious probe setup. Let us
first consider a general setup with a single probe for simplicity.

We will now introduce the implementation of charge conservation
through the system (i.e. in the fictitious probe) which will lead to
the simple result expressed in Eq. (\ref{FCSmu}). We first make the
essential approximation that the inelastic scattering time in the
interacting region is much smaller than
\begin{equation}
\Delta t\equiv h/eV.
\end{equation}
The inverse of the timescale $\Delta t$  is the average rate at
which the external leads inject particles into the system, in the
localized wave-packet terminology \cite{Martin92}. We will call the
scattering processes within time interval $\Delta t$ a `scattering
event'. In this limit of quick scattering we can assume
\emph{sequential} scattering events, as if each $\Delta t$ interval
was an independent few-particle scattering problem, one for each
time
\begin{equation}
t_j\equiv j \Delta t.
\end{equation}
The overlap of the wave-packets which would in principle give
contributions away from the sequential scattering approximation is
assumed to have a negligible effect in the long time limit. Other
works in different contexts \cite{Shelankov03} seem to support this
statement. Furthermore, if one considers small transparency contacts
between the electron source and the fictitious probes, the
sequential scattering approximation is also exact.

The incoming state in each scattering event will be one particle in
each channel of the external leads ($L_1$ and $L_2$ in the setup of
Fig. \ref{fig:setup}), plus a certain state in the probe's leads
$1'$ and $2'$. This state injected from the probe is prepared in a
way so as to compensate for excess charge scattered into the
fictitious probe in all previous events, with the intention of
canceling any current that has flowed into the probe in the past.
The book-keeping of the probe's excess charge is done via an
auxiliary slave degree of freedom $|Q\rangle$ with discrete quantum
numbers $Q=0,\pm 1,\pm 2,\dots$ that count charge transferred to the
probe. The incoming state in leads $1'$ and $2'$ injected by the
probe into the system is a function of $Q$. The time evolution of
the slave state $|Q\rangle$ is constrained so that $Q$ always equals
the total number of electrons that has entered the probe since the
first scattering event. In particular, the time evolution of
$|Q\rangle$ during one scattering event $\Delta t$ is taken to
follow the resulting net charge that was transferred to the probe
during that event. This scheme effectively correlates the initially
uncorrelated scattering events in order to satisfy instantaneous
current conservation through the system, where by instantaneous we
mean at times larger than $\Delta t$ but still much smaller than the
measuring time.

If the incoming state in the  probe's leads is chosen correctly, the
number of $Q$ states between which $|Q\rangle$ will fluctuate during
many scattering events will be bounded, and will be independent of
the total number of events
\begin{equation}
M=eVt/h
\end{equation}
in the total experiment time $t$. This is the underlying principle
of this approach, which will guarantee that the instantaneous charge
fluctuations in the probe will be bounded to a few electrons
throughout the whole measurement process, i.e., the probe current
will be zero and noiseless at frequencies below $eV/\hbar$.
%

The choice that minimizes the charge fluctuations in a single
channel two-legged probe in the absence of superconductors in the
system is the following: if $Q$ at the beginning of the scattering
event is $1$ or $2$, the probe will emit two particles, one through
each 'leg', thereby losing a maximum of $2$ and a minimum of $0$ in
that event; if $Q$ is $0$, $-1$ or $-2$ the probe will not emit any
particle, thereby absorbing a maximum of $2$ and a minimum of $0$.
The resulting fluctuations of $Q$ are bounded in the $[-2,2]$ range.
In some cases, such as the system discussed in the main text, this
range is reduced to $[-1,1]$ since the entangler only emits one
electron of each spin in each scattering event, so that the probe
will never absorb $2$ particles, but a maximum of $1$. The relevance
of this discussion will be apparent in connection with Eq.
(\ref{Wgeneral}), since it will determine the dimensions of the $W$
operator therein.

\subsection{Sequential scattering scheme for the Full Counting
Statistics}

We wish to compute in a general case the characteristic function
\begin{equation}
\chi(\vec{\lambda};M)=\langle e^{i\sum_n\lambda_n
\Delta\hat{N}_n}\rangle=\Tr\left\{\hat{\chi}(\vec{\lambda})\rho(t)\right\}
\label{FCS}
\end{equation}
after a total measuring time interval $t$. Number difference
$\Delta\hat{N}_n\equiv
\hat{N}^\mathrm{out}_{n}-\hat{N}^\mathrm{in}_{n}$ is defined as the
number operator in channel $n$ at time $t$ (scattered outgoing
particle number) minus the number operator at time zero, before any
scattering (incoming particle number). Differentiating $\ln\chi$
respect to the counting fields $\lambda_n$ one obtains the different
transferred charge and current cumulants, Eq. (\ref{Icum}).

Let us include the fictitious probe and expand our Fock space with
the slave degree of freedom $|Q\rangle$. We take the density matrix
of the whole system at time zero equal to
$\rho(0)=\rho^Q(0)\otimes\rho^e(0)$, the second $\rho$ being the
electronic density matrix. As we will see we do not need to specify
the initial state of the slave degree of freedom $\rho^Q(0)$ since
it will not affect our results in the long time limit. The density
matrix is factorized in the localized wave-packet basis
\cite{Martin92},
\begin{equation}
\rho(0)=\rho^Q(0)\otimes\prod_j^\otimes\rho^e_j \label{rho0},
\end{equation}
with the electronic part being
$\rho^e_j\equiv\rho^r_j\otimes\rho^p_j$. Each of these $\rho^e_j$
constitutes the incoming state in each of the $j$ scattering events
corresponding to the time interval $[t_j,t_{j+1}]$. $\rho^r_j$,
which is actually $j$-independent, is the density matrix of the
(uncorrelated in time) electrons coming from the external
reservoirs, and $\rho^p_j$ is the density matrix of the (correlated
in time-through-$Q$) electrons coming from the fictitious probe. As
we mentioned, this matrix $\rho^p_j$ will depend on the state of the
slave degree of freedom $Q$ at the beginning of each scattering
event $j$.

The time evolution from 0 to $t$, $\rho(t)=\hat{U}_t
\rho(0)\hat{U}^+_t$ is split up in the $M$ time intervals of length
$\Delta t$. The sequential scattering approximation amounts to
assuming that in each event each electron group $\rho^e_j$ scatters
completely before the next one does. Therefore
$\hat{U}_t=\hat{U}_{\Delta t}^M$. We defer the discussion on how
$U_{\Delta t}$ operates precisely to a little later.

Since operator $\hat{\chi}(\vec{\lambda})$ will factorize into
contributions for each scattering event, $\hat{\chi}=\prod_j
\hat{\chi}_j$, we can rewrite equation (\ref{FCS}) as
\begin{eqnarray}
\label{chifac}
\chi&=&\Tr_{Q}\left\{\Tr_{M}\left[\hat{\chi}_M U_{\Delta
t}\rho^e_M\Tr_{M-1}\left[\right.\right.\right. \cdots \\
&\cdots&\left.\left.\left.\Tr_{1}\left[\hat{\chi}_1 U_{\Delta
t}\rho^e_1\rho^Q(0)U^+_{\Delta t}\right]U^+_{\Delta
t}\right]U^+_{\Delta t}\cdots\right]\right\}\nn,
\end{eqnarray}
where $\Tr_{j}$ stands for the trace over the $\rho^e_j$ electron
states and $\Tr_Q$ over the $Q$ subspace. An alternative way of
writing this is by induction. Defining an auxiliary operator
$\hat{\Phi}^{(k)}=\sum_{QQ'}|Q\rangle\Phi^{(k)}_{QQ'}\langle Q'|$
such that
\begin{eqnarray}
\hat{\Phi}^{(j)}&=&\Tr_j\left[\hat{\chi}_j U_{\Delta
t}\rho^e_j\hat{\Phi}^{(j-1)}U^+_{\Delta t
}\right]\label{Phidef},\\
\hat{\Phi}^{(0)}&=&\rho^Q(0)\label{Phi0},
\end{eqnarray}
one can see that (\ref{chifac}) and (\ref{FCS}) are equivalent to
\begin{eqnarray}
\label{chiM} \chi(\vec{\lambda};M)&=&\Tr_Q \hat{\Phi}^{(M)}.
\end{eqnarray}

After some algebra, Eq. (\ref{Phidef}) can be recast into the
following sum over the total range of $Q$ values,
\begin{equation}
\Phi^{(j)}_{Q_bQ_b'}=\sum_{Q_aQ_a'}W^{Q_aQ_a'}_{Q_bQ_b'}\Phi^{(j-1)}_{Q_aQ_a'},
\label{Wgeneral}
\end{equation}
with the $W$ superoperator
\begin{equation}
W^{Q_aQ_a'}_{Q_bQ_b'}(\vec{\lambda})=\Tr_j\left[
P_{Q_b'Q_b}\hat{\chi}_j(\vec{\lambda})U_{\Delta t}\rho^e_j
P_{Q_aQ_a'}U^+_{\Delta t}\right] \label{Wexp},
\end{equation}
and $P_{QQ'}\equiv|Q\rangle\langle Q'|$ the generalized projector
within the slave degree of freedom space. We will specify how it
operates in practice a bit later, after Eq. (\ref{Wbar}).

Some words about the meaning of this operator $W$, which is a
central object in this technique, are in order at this point. It is
a superoperator that, for $\vec{\lambda}=0$ simply transforms the
reduced density matrix
$\rho^Q(t_j)=\hat{\Phi}^{(j)}(\vec{\lambda}=0)$ of the slave degree
of freedom at time $t_j$ to the subsequent one
$\hat{\Phi}^{(j+1)}(\vec{\lambda}=0)$ at time $t_{j+1}$. In Eq.
(\ref{Phidef}) we see how $\hat{\Phi}^{(j)}$ is simply
$\hat{\Phi}^{(j-1)}$ to which the incoming state $\rho^e_j$ for
event $j$ is added, is allowed to evolve a time $\Delta t$ (during
which also $\rho^Q$ evolves as dictated by the number of electrons
scattered into the probe), and the scattered electrons are traced
out. The result is the new evolved reduced density matrix for the
slave degree of freedom. For finite $\vec{\lambda}$, the
corresponding counting fields for the scattered electrons are also
included into $\hat{\Phi}^{(j)}$ so as to be able to recover the
desired cumulants of the traced-out electrons after time $t$ from
$\chi=\Tr_Q\hat{\Phi}^{(M)}$. This can be also seen as supplementing
the dynamics of the system with a quantum field term $\propto
\lambda_n$ in the action, in the generalized Keldysh language of
Ref. \onlinecite{Nazarov99}.

By assuming without loss of generality a diagonal initial
$\rho^Q(0)$ and by noting that, by construction, states with
different $Q$ are orthogonal, we can in general take $W$ to be
diagonal
$W^{Q_aQ_a'}_{Q_bQ_b'}=\delta_{Q_aQ_a'}\delta_{Q_bQ_b'}\overline{W}_{Q_bQ_a}$,
and $\Phi^{(j)}_{QQ'}=\delta_{QQ'}\Phi^{(j)}_{Q}$. Physically this
means that sequentially taking out of the system the scattered
electrons (tracing them out) forbids the $Q$ counter to remain in a
coherent superposition, since the electron that generated it has
been `measured'. Therefore (\ref{chiM}) finally becomes
\begin{equation}
\chi(\vec{\lambda};M)=\sum_{Q_aQ_b}\overline{W}^M_{Q_bQ_a}\rho^Q_{Q_aQ_a}(0)
\end{equation}
(note the $M^\mathrm{th}$ power of the $\overline{W}$ matrix). The
following alternative and useful form for (\ref{Wexp}) can be
obtained by writing $|Q\rangle\langle Q|\rho^e_j|Q\rangle\langle
Q|=|\phi^e Q\rangle\langle\phi^e Q|$, in the case of a pure incoming
state in the external leads,
\begin{equation}
\overline{W}_{Q_bQ_a}(\vec{\lambda})=\langle \phi^e Q_a|U^+_{\Delta
t} P_{Q_bQ_b}\hat{\chi}_j(\vec{\lambda})U_{\Delta t} |\phi^e
Q_a\rangle \label{Wbar},
\end{equation}
where $|\phi^e Q\rangle$ stands now for the incoming electronic state (through
all leads) that corresponds to a given value $Q$ of the slave degree of
freedom.

Let us analyze the action of the evolution operator $U_{\Delta t}$
in the above equation. Since we assume that particles scatter fully
in time $\Delta t$, the action of $U_{\Delta t}$ on the electrons is
written in terms of the global scattering matrix $b^+_n=U_{\Delta
t}a^+_n U^+_{\Delta t}=\sum_m S_{nm}a^+_m$, where $a^+_n$ are the
electron creation operators in the different leads (including
fictitious ones) of the system \footnote{Note the difference with
the notation in \cite{Blanter00}. When there is time reversal
symmetry both choices are equivalent.}. The effect of $U_{\Delta t}$
on the $\hat{Q}$ degree of freedom is merely to update it with the
net number of electrons scattered into the fictitious leads, fixing
$\hat{Q}_{t_{j+1}}-\hat{Q}_{t_j}=\Delta \hat{N}_p$, where
$\Delta\hat{N}_p$ is the number of electrons absorbed by the probe
in the event. This implies that $P_{Q_bQ_b}$ in Eq. (\ref{Wbar}),
which projects on the subspace with $Q=Q_b$, can be substituted by
the electron-only operator that projects over scattered electronic
states that satisfy
$\hat{N}^\mathrm{out}_p=\hat{N}^\mathrm{in}_p+Q_b-Q_a$, where
$\hat{N}^\mathrm{out}_p$ is the number operator for fermions
scattered into the probe, $\hat{N}^\mathrm{in}_p$ is the number of
electrons incident from the probe into the system at the beginning
of the scattering event, and $Q_a$ is the value of $Q$ also at the
beginning of the scattering event.

As anticipated just before the beginning of this subsection, the
value of $\hat{N}^\mathrm{in}_p$ on $|\phi^e Q_a\rangle$ is a
function of $Q_b$, and should be chosen properly so as to compensate
for a given excess probe charge $Q_a$ at the beginning of a given
scattering event. That way the fluctuations of the probe's excess
charge $Q$ will be minimum, although the precise choice does not
affect the result as long as the resulting range of fluctuations of
$Q$ does not scale with measurement time $t$. As already discussed,
for most cases the optimum choice is $N^{in}_p(Q)=g(Q)$, with
\begin{eqnarray}
g(1)&=&2~~~ \textrm{(one electron in each lead of the probe)},\nn\\
g(0)&=&g(-1)=0, \label{choice}
\end{eqnarray}
which gives $Q\in[-1,1]$, and a $3\times 3$ $\overline{W}$ matrix.

To finish with the discussion of Eq. (\ref{Wbar}), recall that
$\hat{\chi_j}=e^{i\sum_n\lambda_n(\hat{N}^\mathrm{out}_n-\hat{N}^\mathrm{in}_n)}$
and that a useful relation for the case of a single channel mode $n$
in which the eigenvalues of $\hat{N}_n$ are zero and one is
$e^{i\lambda_n \hat{N}_n}=1+(e^{i\lambda_n}-1)\hat{N}_n$.

The whole  Levitov-Lee-Lesovik formulation of FCS \cite{Levitov96}
is well defined only in the long time limit. In such limit it is
clear that expression (\ref{Wbar}) is dominated by the biggest
eigenvalues $\mumax$ of $\overline{W}$. All of its eigenvalues
satisfy $|\mu|\le 1$ for real values of $\vec{\lambda}$, so that
those that are not close to 1 for small values of $\vec{\lambda}$
(around which we take derivatives to compute cumulants) will
exponentiate to zero when $M\rightarrow \infty$. In all cases we
examined only one eigenvalue $\mumax$ would not exponentiate to
zero, although it can have finite degeneracy. In general, we have
the following asymptotic property, valid for any degeneracy of
$\mumax$,
\begin{equation}
\ln\left[\chi(\vec{\lambda};M)\right]=M \ln\left[\mumax\right] +
\mathcal{O}(1) \label{FCSmu}.
\end{equation}
We can define a new generating function
\begin{equation}
\ln\chi_I(\vec{\lambda})=\lim_{M\rightarrow\infty}\frac{\ln\chi(\vec{\lambda};M)}{M}.
\label{ChiIDef}
\end{equation}
It can be shown that this function generates the zero frequency
limit of current cumulants
\begin{equation}
\langle\langle
I_n(\omega=0)^k\rangle\rangle=\frac{e^{k+1}|V|}{h}(-i)^k\left.\partial^k_{\lambda_n}\ln\chi_I\right|_{\vec{\lambda}=0},
\label{Icum}
\end{equation}
being $e$ here the electron charge and $k$ the order of the
cumulant, $k=1$ for the average current, $k=2$ for the shot noise,
and so on.

We can identify
\begin{equation}
\chi_I(\vec{\lambda})=\mumax(\vec{\lambda}) \label{chiI}.
\end{equation}
This is our final result. $\mumax$ is the eigenvalue of Eq.
(\ref{Wbar}) that equals $1$ when all counting fields $\lambda_n$
are taken to zero.

The generalization to multiple probes is very straightforward. Given
the optimum choice of Eq. (\ref{choice}), the solution of an
$N-$probe setup will involve the diagonalization of an $3^N\times
3^N$ $\overline{W}$ matrix similar to Eq. (\ref{Wbar}) where $Q$ is
changed to $\vec{Q}$, a vector of the $N$ corresponding slave
degrees of freedom. On the other hand, to implement charge
conservation in probes with more than one channel per leg (or more
than two legs), such as non-spin-conserving probes, the formalism
would require a slightly different expression for Eq. (\ref{choice})
and a consequently bigger dimension for $\overline{W}$, but would
otherwise remain quite the same.

A summary of the above results is given in Sec. \ref{sec:technique}.

We have successfully compared the present method to Langevin
techniques in scenarios where the latter is directly applicable
(uncorrelated spins), obtaining identical results in all cases. Some
simple examples are the FCS of a single channel wire with contact
transmissions $T_1$ (see Appendix \ref{ap:comp} for the details),
the case of a Mach-Zehnder interferometer or an NS junction, for
which both this and the Langevin method \cite{Beenakker92} yield
identical results for $\chi_I$. We would like to mention that even
in the presence of correlated spins the Langevin technique can be
extended to include correlation effects to some extent, as is
discussed in Appendix \ref{ap:comp}.

\section{Comparison of the method to previous techniques \label{ap:comp}}

In this appendix we will show with a simple example how the proposed
method yields identical results to the ones obtained with previous
Langevin technique (applicable in such case), which we generalize
here to yield the FCS, instead of individual current cumulants. We
also sketch how a Langevin derivation of the beam splitter FCS could
be attempted by extending the technique, and a comparison to our
results. The purpose of this section is twofold. First we wish to
make a convincing case that our method actually recovers known
results, but goes beyond them in other cases, and secondly, that it
indeed yields the FCS in the presence of an inelastic probe, and not
merely a dephasing probe as could be thought from the unusual real
time sequential scattering picture.

\begin{figure}
\includegraphics[width=6cm]{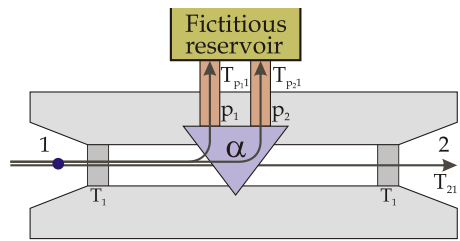}
\caption{(Color online) Single channel wire geometry with fictitious
probe and non-transparent contacts.} \label{fig:setup_ap}
\end{figure}

We will first do our comparison in the possibly simplest system one
can think of, a zero-temperature single channel conductor for
spinless fermions, see Fig. \ref{fig:setup_ap}. We will assume
symmetric contacts to the (real) reservoirs with transmission $T_1$.
A fictitious inelastic probe will be connected between the two
contacts with transmission amplitude $\alpha$, and scattering matrix
(\ref{Salpha}).

As discussed in detail in Ref. \onlinecite{Blanter00}, within the
Langevin approach, the current fluctuations in the presence of the
probe should be corrected by the feedback due to the instantaneous
fluctuations of the probes voltage, which react to cancel any
current flowing into the probe. Thus the current fluctuation flowing
into the right reservoir reads
\begin{equation}
\Delta I_2=\delta I_2+\frac{T_{2p}}{T_{1p}+T_{2p}}\left(\delta
I_{p_1}+\delta I_{p_2}\right),
\end{equation}
where $\delta I_2$ correspond to the current fluctuations flowing
out of the system through lead $2$ with a static potential in the
probe, whereas $\delta I_{p_1}$ and $\delta I_{p_2}$ are the
analogous currents leaving the system through legs $p_1$ and $p_2$
of the probe. $T_{nm}=T_{mn}$ are the transmission probabilities
between channels $m$ and $n$. We make use of the compact notation
$T_{np}=T_{n{p_1}}+T_{n{p_1}}$. The static potential in the probe
for $\delta I_n$ is chosen so as to cancel any average current into
the probe.

In our case, left-right symmetry implies
\begin{equation}
\Delta I_2=\delta I_2+\frac{1}{2}\left(\delta I_{p_1}+\delta
I_{p_2}\right) \label{LangevinDI}.
\end{equation}
At this point, what one usually finds in the literature is a
calculation of cumulants of certain order. It is possible however to
compute them all at once and recover the FCS, as we show in the
following. Define the characteristic function with a static
potential $\mu_p=eV/2$ in the probe (the value which gives average
current conservation) and with \emph{two} counting fields, one
($\lambda_2$) that counts particles flowing into the rightmost
reservoir, and another ($\lambda_p$) that counts particles scattered
into the probe
\begin{equation}
\chi(\lambda_2,\lambda_p)=\langle e^{i\lambda_2 \Delta
N_2+i\lambda_p(\Delta N_{p_1}+\Delta N_{p_2})}\rangle.
\label{chiLang}
\end{equation}

Since at this point there are still no probe fluctuations (each
energy is independent from the rest), one can write
$\chi(\lambda_2,\lambda_p)$ as the product of two characteristic
functions, one for particles in the interval $[0,\mu_p]$ and another
in the $[\mu_p,eV]$\cite{Levitov96}. We have
\begin{eqnarray}
\chi(\lambda_2,\lambda_p)&=&\langle
1|\hat{\chi}_2(\lambda_2)\hat{\chi}_{p_1}(\lambda_p)\hat{\chi}_{p_2}(\lambda_p)|1\rangle^{M/2}\nn\\
&&\langle
0|\hat{\chi}_2(\lambda_2)\hat{\chi}_{p_1}(\lambda_p)\hat{\chi}_{p_2}(\lambda_p)|0\rangle^{M/2},
\end{eqnarray}
where $\hat{\chi}_n(\lambda)\equiv
1+\left(e^{i\lambda}-1\right)\hat{N}_n$, $|1\rangle$ is the
scattered state at an energy below $\mu_p$ (i.e., with a full state
coming from the probe), and $|0\rangle$ is a state above $\mu_p$
(empty probe). Working out the algebra we get for the \emph{current}
characteristic function, Eq. (\ref{ChiIDef}),
\begin{eqnarray}
\chi_I(\lambda_2,\lambda_p)&=&\left\{\left[1+(e^{i\lambda_2}-1)T_{12}+(e^{i\lambda_2-i\lambda_p}-1)T_{2p}\right]\right.\nn\\
&&\times\left.\left[1+(e^{i\lambda_2}-1)T_{12}+(e^{i\lambda_p}-1)T_{2p}\right]\right\}^{\frac{1}{2}}\nn,
\end{eqnarray}
where $T_{12}=T_1^2(1-\alpha)/[2-\alpha-T_1(1-\alpha)]^2$ and
$T_{2p}=T_1\alpha/[2-\alpha-T_1(1-\alpha)]$.

To include the self-consistent voltage fluctuations of the probe, we
return to Eq. (\ref{LangevinDI}). It is easy to see that the
function
\begin{equation}
\chi_I(\lambda_2)\equiv\chi_I(\lambda_2,\frac{\lambda_2}{2})=1+(e^{i\lambda_2}-1)T_{12}+(e^{i\lambda_2/2}-1)T_{2p}
\label{LangevinChi}
\end{equation}
generates the cumulants of $\Delta I_2$, instead of $\delta I_2$,
and therefore is the proper FCS solution of the Langevin approach.

On the other hand, the method we have developed involves, in this
simple system and for the same choice of $g(Q)$ as in Eq.
(\ref{choice}), the following expression of the $\overline{W}$
matrix in (\ref{Wbar})
\begin{eqnarray}
\overline{W}= \left(\begin{array}{ccc}
a&0&0\\
c&a&b\\
0&c&a
\end{array}\right),
\end{eqnarray}
with $a=1+(e^{i\lambda_2}-1)T_{12}-T_{2p}$, $b=e^{i\lambda_2}T_{2p}$
and $c=T_{2p}$. The highest eigenvalue of this matrix is
$a+\sqrt{bc}$, which indeed equals the Langevin result
(\ref{LangevinChi}). 

This example clarifies the fact that the fictitious probe we are
describing within our approach is inelastic since, as is evident
within the Langevin approach, a particle scattered into the probe at
a certain energy can abandon it at any other energy in the interval
$[0,eV]$. In particular, note that the the current through the
system when $T_1=1$ and $\alpha=1$ is noiseless, i.e., the Fano
factor as derived from Eq. (\ref{LangevinChi}) is $F=0$, as opposed
to $F=1/4$ that would result from the quasi-elastic probe
\cite{Blanter00}.

\subsection{Langevin technique with spin correlations}

The Langevin language makes use of one crucial assumption, that the
currents flowing into the system are those which result from some
static chemical potentials in the real non-interacting reservoirs,
and which are therefore spin-uncorrelated. Can one use it to compute
current fluctuations when the electrons injected into the system are
in a tailored spin state, such as non-local spin singlets arriving
simultaneously on the beam splitter in the main text (Fig.
\ref{fig:setup})? The answer is "no", but one can actually go quite
far in this direction. One can modify the above scheme to try to
account for the peculiar spin correlations, although it only works
up to the second cumulant, deviations appearing from the third
cumulant onward. This works as follows. One could compute the
equivalent chemical potential for spin-up and spin-down electrons in
each of the incoming leads as if they were completely uncorrelated,
but then try to preserve the correlation information by inserting
the proper spin-correlated state in Eq. (\ref{chiLang}). It is
rather unclear in this case whether one is thereby assuming that
electrons arriving at the same time into the two leads $L_1$ and
$L_2$ are spin entangled, or whether it is electrons with definite
and equal energy which are non-locally spin entangled. Remarkably,
the result for the shot noise agrees with the one obtained with our
time-resolved technique, where these questions are fully under
control, but higher cumulants do not. From a mathematical point of
view it is hardly surprising that the FCS from both techniques does
not agree in the general case, since through Langevin one obtains an
explicit form of the generating function in terms of elementary
functions, while with the time-resolved technique the latter is the
solution of a non-reducible ninth order polynomial, Eq.
(\ref{chidet}), which is known not to have a closed form in general.
The discrepancy between the methods, which is rather small in most
cases, is most likely due to the fact that in the Langevin language
it is not possible to encode the time-resolved spin correlation
information of the incoming current.

\begin{figure}
\includegraphics[width=7cm]{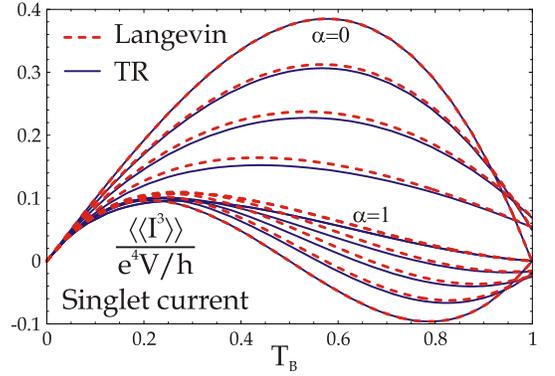}
\caption{(Color online) The results for the singlet current skewness
in the beam splitter geometry calculated within the modified
Langevin approach begin to deviate slightly from those of the
time-resolved (TR) technique, Fig. \ref{fig:skewness}. Note that the
deviation is most apparent in the intermediate $\alpha$ limit. Shot
noise results are identical. At this level, however, both techniques
are qualitatively equivalent.} \label{fig:comp}
\end{figure}

We now sketch in more detail how the attempt at calculating the FCS
of the beam splitter within the Langevin language should go. All
chemical potentials are in principle spin-dependent, and
transmissions (for $\theta\neq 0$) will connect different spins. The
equivalent chemical potential in leads $L_1$ and $L_2$ is
$\mu_{L}^\sigma=eV/2$, since only one out of two particles has, say,
spin-up. The average chemical potential in the voltage probe reads
\begin{equation}
\mu_p^\sigma=\frac{eV}{2}\frac{\sum_{\sigma'}T_{pL}^{\sigma\sigma'}}{\sum_{\sigma'}T_{pL}^{\sigma\sigma'}+T_{pR}^{\sigma\sigma'}}=
\frac{eV}{2}\frac{2-T_B(1-\alpha)-\alpha}{2-\alpha(1-T_B)},
\end{equation}
where
$T_{pL}^{\sigma\sigma'}=T_{p_1L_1}^{\sigma\sigma'}+T_{p_2L_1}^{\sigma\sigma'}+T_{p_1L_2}^{\sigma\sigma'}+T_{p_2L_2}^{\sigma\sigma'}$,
and $T_{p_1L_1}^{\sigma\sigma'}$ is the transmission probability
from lead $L_1$ and spin $\sigma'$ to probe leg $p_1$ and spin
$\sigma$. Equivalently for $T_{pR}^{\sigma\sigma'}$. The full
current fluctuation flowing into the right reservoir through lead
$R_1$ reads
\begin{eqnarray}
\Delta I_{R_1}&=&\sum_{\sigma}\Delta
I_{R_1}^\sigma=\sum_{\sigma}\left[\delta
I_{R_1}^{\sigma}+\sum_{\sigma_2}\frac{T_{pR_1}^{\sigma\sigma_2}(\delta
I_{p_1}^\sigma+\delta
I_{p_2}^\sigma)}{\sum_{\sigma_1}T_{pL}^{\sigma_2\sigma_1}+T_{pR}^{\sigma_2\sigma_1}}\right]\nn\\
&=&\sum_{\sigma}\left[\delta
I_{R_1}^{\sigma}+\frac{T_B(1-T)}{2-\alpha(1-T_B)}(\delta
I_{p_1}^\sigma+\delta I_{p_2}^\sigma)\right],
\end{eqnarray}
so that the substitution
\begin{eqnarray}
\lambda_{R_1}^\uparrow&=&\lambda_{R_1}^\downarrow=\lambda, \\
\lambda_{p}^\uparrow&=&\lambda_{p}^\downarrow=\frac{T_B(1-T)}{2-\alpha(1-T_B)}\lambda
\end{eqnarray}
into
\begin{widetext}
\begin{eqnarray*}
\chi(\lambda_{R_1}^\uparrow,\lambda_{R_1}^\downarrow,\lambda_p^\uparrow,\lambda_p^\downarrow)=\langle
1|\prod_\sigma\hat{\chi}_{R_1}(\lambda_{R_1}^{\sigma})\hat{\chi}_{p_1}(\lambda_p^\sigma)\hat{\chi}_{p_2}^\sigma(\lambda_p^\sigma)|1\rangle^{\frac{\mu_p}{eV}}\times\langle
0|\prod_\sigma\hat{\chi}_{R_1}(\lambda_{R_1}^{\sigma})\hat{\chi}_{p_1}(\lambda_p^\sigma)\hat{\chi}_{p_2}^\sigma(\lambda_p^\sigma)|0\rangle^{\frac{1-\mu_p}{eV}}
\end{eqnarray*}
\end{widetext}
yields the characteristic function for current fluctuations flowing
out into the right reservoir through the detector in lead $R_1$. Now
state $|0\rangle$ (state $|1\rangle$) is the scattered state
corresponding to the proper spin-correlated pair coming into leads
$L_1$ and $L_2$, cf. Eq. (\ref{states}), together with an empty
(full) state coming from the fictitious probe.

If one inserts the resulting solution for $\chi_I=\chi^{1/M}$ into
Eq. (\ref{chidet}), one indeed obtains zero to order $(i\lambda)^2$,
but not to higher orders in $i\lambda$. This indeed implies that up
to second order cumulants both methods agree (again confirming the
statement that we are modelling inelastic and not elastic
scattering), but not beyond. A comparison of the third cumulant is
illustrated in Fig. \ref{fig:comp}, where one can appreciate the
small deviation. The results at this level remain qualitatively
equivalent, however.

\bibliography{BeamSplitterPRB}
\end{document}